\documentclass[aps,pra,twocolumn,showpacs]{revtex4}
\usepackage[pdfview=FitV,pdfstartview=FitV]{hyperref}
\usepackage{amsmath,amssymb,amsfonts,empheq,bm}
\usepackage{graphicx}
\usepackage{epsfig}
\usepackage{color}

\newcommand{\ii}{{i}}

\begin{document}

\title{Ring model for trapped condensates with synthetic spin-orbit coupling}

\author{Xing Chen}
\affiliation{Beijing National Laboratory for Condensed Matter Physics,Institute of Physics, Chinese Academy of Sciences, Beijing 100190, China}
\author{Michael Rabinovic}
\affiliation{Laboratoire Kastler-Brossel, Ecole Normale Superieure, 24 rue Lhomond, 75005 Paris, France}
\author{Brandon M. Anderson}
\affiliation{Joint Quantum Institute, National Institute of Standards and
Technology and University of Maryland, Gaithersburg, MD 20899, USA}
\author{Luis Santos}
\affiliation{Institut f\"ur Theoretische Physik, Leibniz Universit\"at Hannover, Appelstr. 2, DE-30167 Hannover, Germany}

\begin{abstract}
We derive an effective ring model in momentum space for trapped bosons with synthetic spin-orbit coupling. This effective model is characterized
by a peculiar form of the inter particle interactions, which is crucially modified by the external confinement. The ring model allows for an intuitive understanding of the
phase diagram of trapped condensates with isotropic spin-orbit coupling, and in particular for the existence of skyrmion lattice phases.
The model, which may be generally applied for spinor condensates of arbitrary spin and spin-dependent interactions, is illustrated
for the particular cases of spin-$1/2$ and spin-$1$ condensates.
\end{abstract}


\maketitle



\section{Introduction}
\label{sec:Introduction}

Synthetic electromagnetism in ultra cold neutral gases has attracted great
interest~\cite{Dalibard2011,Goldman2013} in recent years. In spite of the absence of charge, the use of appropriate laser arrangements has allowed for mimicking the
effect of artificial magnetic fields both in the continuum~\cite{Lin2009} and in
optical lattices~\cite{Aidelsburger2013,Miyake2013}. Moreover, the internal level structure of the atoms may be employed to create
synthetic spin-orbit coupling~(SOC)~\cite{Lin2011,Wang2012,Cheuk2012}, an essential ingredient in many condensed-matter phenomena.

The physics of degenerate quantum gases in the presence of SOC has attracted a large deal of theoretical
attention~(for recent reviews see Refs.~\cite{Zhai2012,Galitski2013,Zhou2013b,Zhai2014} and references therein). A particular emphasis has been paid
to the case of an equal admixture of Rashba and Dresselhaus SOC, since this is the situation that has been experimentally
realized up to now~\cite{Lin2011,Wang2012,Cheuk2012}. The physics of Bose Einstein Condensates (BEC)  in the presence of isotropic SOC,
such as Rashba or Dresselhaus, is however
particularly interesting due to the associated peculiar ring-like dispersion. In the homogeneous case~(in the absence of a trap) the mean-field ground state
of a two-dimensional spin-$1/2$ BEC breaks polar symmetry spontaneously being characterized by the so-called plane-wave or stripe phases, respectively corresponding
to one peak or two opposite momentum peaks in the dispersion ring~\cite{Wang2010}. The presence of a harmonic trap may significantly enrich the
ground-state phase diagram, leading to the presence of half quantum vortex phases~\cite{Stanescu2008,Wu2011,Ramachandhran2012} and
skyrmion lattice patterns~\cite{Sinha2011,Hu2012}. The effects of SOC for the case of BECs with higher spin have been also
discussed~\cite{Zhai2012,Zhai2014,Xu2012}. In particular, a spin-$1$ BEC with SOC~(which could be generated
using pulsed magnetic fields~\cite{Anderson2013,Xu2013}) may present triangular and square skyrmion lattice phases~\cite{Xu2012,Ruokokoski2012}.

In this paper we provide a simplified picture that allows for an intuitive understanding of the
physics behind the various ground-state phases of trapped BECs in the presence of Rashba (or Dresselhaus) SOC. By exploiting the ring-like
form of the dispersion, we derive an effective quasi-one-dimensional model in momentum space. As for the homogeneous case~\cite{Gopalakrishnan2011,Zhou2013}
the effective quasi-1D model is characterized by two types of interaction, an effective long-range interaction in momentum space, and a destruction/creation
of pairs of atoms with opposite momentum on the Rashba ring. We show, however, that the presence of the trap crucially modifies the
form of the interactions, and that this trap-induced modification of the interactions in the effective quasi-1D model explains the numerically observed skyrmion lattice phases of different geometries~\cite{Sinha2011,Hu2012,Xu2012}.

The structure of the paper is as follows. In Sec.~\ref{sec:2D} we introduce the two-dimensional model of spin-$1/2$ BECs with isotropic SOC.
Section~\ref{sec:RingModelSpin12} discusses the derivation of the effective ring model for spin-$1/2$ BECs, showing that the quasi-1D model allows for
an intuitive understanding of the ground-state phase diagram. In Sec.~\ref{sec:RingModelSpin1} we illustrate the general use of the ring model
with a discussion of spin-$1$ BECs. Finally in Sec.~\ref{sec:Outlook} we summarize and comment on further applications.


\section{Two-dimensional condensates with spin-orbit coupling}
\label{sec:2D}

We consider in the following a trapped two-dimensional pseudo-spin-F BEC with spin-independent interactions in the presence of an isotropic synthetic SOC.
The condensate is described by the energy functional
$E=E_{\mathrm{SOC}}+E_{\mathrm{T}}+E_{\mathrm{I}}$ ,
where
\begin{eqnarray}
 E_{\mathrm{SOC}}[\boldsymbol\Psi]&=&\frac{1}{2m} \int \mathrm{d}^{2}\vec{r}\, \boldsymbol\Psi^{\dagger} \left(-\ii \hbar\vec\nabla\!-\!\hbar\frac{\kappa}{F}  \vec{\bf F}_\bot \right)^{2}\boldsymbol\Psi,  \label{eq:HSOC}
\\
 E_{\mathrm{T}}[\boldsymbol\Psi]&=&\int \mathrm{d}^{2}\vec{r}\, V(r) \boldsymbol\Psi^{\dagger}\cdot \boldsymbol\Psi,  \label{eq:HT}
\\
 E_{\mathrm{I}}[\boldsymbol\Psi]&=& \frac{g}{2} \int \mathrm{d}^{2}\vec{r}\, \left ( \boldsymbol\Psi^{\dagger}\cdot \boldsymbol\Psi \right )^2,
 \label{eq:HI}
\end{eqnarray}
characterize, respectively, the spin-orbit coupling term, the trap energy, and the interaction energy. In the previous expressions the momentum $\kappa$ characterizes the SOC strength,
$V(r)=m\omega^{2}r^{2}/2$ is the isotropic harmonic trap on the $xy$ plane. Without loss of generality, we chose the spin-orbit coupling vector $\mathbf{F}_\bot = \mathbf{F}_x \vec{e}_x + \mathbf{F}_y \vec{e}_y$ to be the in-plane component of the spin vector with components $[{\bf F}_a, {\bf F}_b] = \ii \epsilon_{abc} {\bf F}_c$. Note that Dresselhaus or Rashba forms
will provide identical results, up to an unitary rotation.  In the previous equations,
$ \boldsymbol\Psi(\vec r)$ is the two-component spinor wave function. Note that we are hence performing a mean-field analysis, although the ring model discussed below
may be used as well beyond the mean-field approximation.
The condensate physics is hence given by the 2D Gross-Pitaevskii equation~(GPE):
\begin{equation}
\ii \hbar\frac{\partial}{\partial t}\boldsymbol\Psi=\left [ \frac{\left(-\ii \hbar\vec\nabla\!-\!\hbar \frac{\kappa}{F}  \vec{\mathbf{F}}_\bot \right)^{2}}{2m}+V(r)+g\left ( \boldsymbol\Psi^{\dagger}\cdot \boldsymbol\Psi \right )\right ] \boldsymbol\Psi
\label{eq:GPE2D}
\end{equation}

 In the following we assume a dominant SOC, i.e. $\hbar^2\kappa^2/2m\gg\hbar\omega$. We also consider that $\hbar\omega$ is much greater than the interaction energy per particle.
The latter condition leads in absence of SOC to a Gaussian BEC in the ground-state of the harmonic trap. The situation is radically
different in the presence of SOC, where for weak interactions the system presents a series of phases and phase transitions. For spin-1/2, these include two
half-vortex phases (HV(1/2) and HV(3/2)) and a skyrmion lattice phase~\cite{Stanescu2008,Wu2011,Ramachandhran2012,Sinha2011,Hu2012}~(for larger interactions the system enters in the so-called stripe or plane-wave phase~\cite{Wang2010}). Whereas the physics behind the half-vortex phases is quite clear,
the energetic justification of the skyrmion lattice phase is on the contrary not well understood.
We develop below a simplified ring model that will allow us for an intuitive understanding of the appearance of the lattice phase.

\section{Ring model for spin-1/2 condensates}
\label{sec:RingModelSpin12}

\subsection{Projection on the lowest energy branch}
\label{subsec:Projection}

We now consider the case of $F=1/2$. The condensate is best described in momentum space, $$\boldsymbol\Psi(\vec r )=\int \frac{d^2k}{(2\pi)^2} e^{\ii \vec k \cdot \vec r}  \tilde{\boldsymbol\Psi}(\vec k ),$$ with
$\vec k=(k,\phi)$ in polar coordinates.
The spin-orbit part of the energy functional, $$E_{\mathrm{SOC}}=\int \frac{d^2 k}{(2\pi)^2}\tilde{\boldsymbol\Psi}^\dag(\vec k) \frac{\hbar^2(\vec k -\kappa \vec{\boldsymbol\sigma_\bot})^2}{2m}\tilde{\boldsymbol\Psi}(\vec k),$$ presents two eigenenergy branches,
$$\epsilon_\pm(k)=\hbar^2(k\pm\kappa)^2/2m.$$
(The in-plane vector of Pauli matrices is $\vec{\bf F}_\bot / F = \vec{\boldsymbol\sigma}_\bot = {\boldsymbol\sigma}_x \vec{e}_x + {\boldsymbol\sigma}_y \vec{e}_y$). Due to the dominant SOC the BEC physics may be restricted to the lowest
branch, $\epsilon_-(k)$, which is characterized by the eigenvector
$${ \boldsymbol\eta}_{-}(\phi)=\frac{1}{\sqrt{2}}\binom{e^{\mathrm{-i}\phi}}{1}.$$ The spinor acquires hence the form
$\tilde{ \boldsymbol\Psi}(\vec k)=\psi(\vec k) \boldsymbol\eta_-(\phi)$. Note that $\epsilon_-(k)$ has a mexican-hat form. For a dominant SOC the BEC
occupies the momentum space region around the ring-like dispersion minimum~(Rashba ring).

\subsection{Trap energy}
\label{subsec:TrapEnergy}

In absence of trapping the Bose gas condenses at one or more points of the classical minimum of the Rashba ring.~\cite{Wang2010,Stanescu2008}.
The harmonic trapping introduces an effective radial and angular dispersion in momentum space:
\begin{equation}
E_{\mathrm{T}}=\int \frac{d^2 k}{(2\pi)^2} \tilde{\boldsymbol\Psi}^\dag(\vec k)\left (\frac{-m\omega^2}{2}\nabla_{\vec k}^2 \right )\tilde{\boldsymbol \Psi}(\vec k).
\end{equation}
Due to the polar symmetry, and for a dominant SOC, we may introduce the separation of coordinates $\psi(\vec k)\simeq G(\phi)f(k)/\sqrt{k}$.
The radial part, $f(k)$, obeys the 1D Hamiltonian $-\frac{m\omega^2}{2} \partial_k^2 +\frac{\hbar^2}{2m}\left (k-\kappa\right )^2$,
characterized by a harmonic energy spectrum $\hbar\omega (n+1/2)$. Since $\hbar\omega$ is  much larger than the interaction energy
we may consider that only $n=0$ is populated, and hence $f(k)=A e ^{-(k-\kappa)^2l_0^2/2}$, where $l_0=\sqrt{\hbar/m\omega}$ is the oscillator length, and
$A$ is a normalization constant that we determine below.

The physics of the weakly interacting Bose gas in this approximation is characterized entirely by the angular dependent $G(\phi)$.
For a dominant SOC we may approximate $k^{-2}\simeq \kappa^{-2}$, and re-write:
\begin{equation}
E_{\mathrm{SOC}}+E_{\mathrm{T}}=\frac{m\omega^2}{2\kappa^2}  \int d\phi\, G(\phi)^*
\left ( \hat l_z- \frac{1}{2}\right )^2 G(\phi),
\label{eq:HSOCT-sim}
\end{equation}
with $\hat l_z=-\ii \partial_\phi$ the angular momentum around the $z$ axis. Note that the shift in the angular dispersion of $1/2$ results from the Berry's phase of $\pi$ that arises from encircling the Rashba ring. This cannot be eliminated by a gauge transformation without inducing twisted boundary conditions in $G(\phi)$.
We now impose the normalization $\int dk |f(k)|^2=(2\pi)^2$ and $\int d\phi |G(\phi)|^2=1$, which results
in $\int d^2 r \boldsymbol\Psi(\vec r)^\dag\cdot\boldsymbol\Psi(\vec r)=1$. This fixes the normalization constant $A$. We obtain in this way the final
form of the spinor in momentum space:
\begin{equation}
\tilde{\boldsymbol\Psi}(\vec k)=2\pi^{3/4}\sqrt{\frac{l_0}{k}}e^{-l_0^2(k-\kappa)^2/2}G(\phi) { \boldsymbol\eta}_{-}(\phi).
\label{eq:Psi}
\end{equation}


\subsection{Interaction energy}
\label{subsec:InteractionEnergy}

In order to evaluate the interaction energy, it is convenient to re-express the spinor wavefunction in coordinate space. To this aim we first decompose $G(\phi)$ into the different angular momentum components,
$$G(\phi)=\sum_l a_l e^{\ii l\phi}/\sqrt{2\pi},$$ with $\sum_l |a_l|^2=1$.
We may employ the approximate identity
\begin{equation}
\int dq \sqrt{q}e^{-(q-\tilde\kappa)^2/2} J_l(q s)\simeq \sqrt{2\pi\tilde\kappa}e^{-s^2/2} J_l(\tilde \kappa s),
\end{equation}
with $s \equiv r/l_0$ and $J_l$ the Bessel function of first kind. Introducing the dimensionless parameter $\tilde\kappa\equiv\kappa l_0$, the precious identity requires $\tilde\kappa\gg l$. The latter implies that the angular wavefunction $G(\phi)$ must have
a sufficiently large  angular spread, such that single-particle energy satisfies $E_{\mathrm{SOC}}+E_{\mathrm{T}}\ll \hbar\omega$~(\emph{thin ring limit}).
In what follows, we assume the thin-ring limit unless otherwise stated. This assumption is the key assumption in the development of the ring model below.
Using the previous identity, we may easily obtain the form of the spinor in coordinate space~($\vec r=(r,\alpha)$):
\begin{equation}
\boldsymbol\Psi(\vec  r)=\frac{\sqrt{\tilde \kappa}\, e^{-s^2/2}}{l_0\sqrt{2\sqrt{\pi}}}  \sum _l a_l
\binom
{\ii^{l-1} e^{\ii (l-1)\alpha}J_{l-1}(\tilde\kappa s)}
{\ii^{l}e^{\ii l\alpha}J_{l}(\tilde\kappa s)}.
\label{eq:spinor}
\end{equation}

\begin{figure}[t]
\includegraphics[width=\columnwidth]{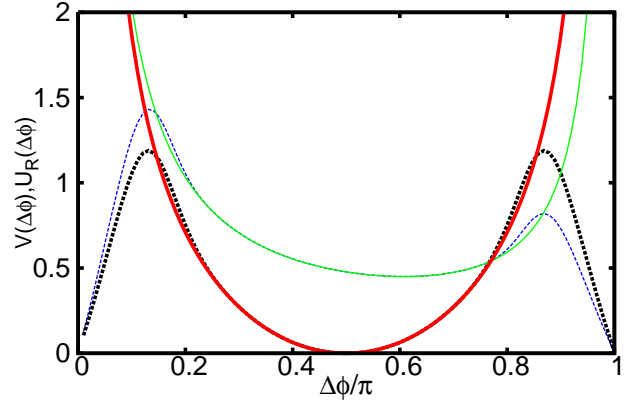}
\caption{(Color online) Functions $V(\Delta\phi)$~(dashed curve) and $U_R(\Delta\phi)\equiv\Re(U(\Delta\phi))$~(bold dashed curve) for a spin-$1/2$ BEC with $\tilde\kappa=20$.
The final expressions are of the form $V(\Delta\phi)=V_0(\Delta\phi)f_0(\Delta\phi)+0.59 V_0(\pi-\Delta\phi)f_\pi(\Delta\phi)
+V_{a}(\Delta\phi)[1-f_0(\Delta\phi)-f_\pi(\Delta\phi)]$, where $f_0(\phi)=e^{-(\phi/0.5\pi)^4}$ and $f_\pi(\phi)=e^{-((\pi-\phi)/0.5\pi)^4}$ are interpolating functions.
Similarly for $U$, $\Re[U(\Delta\phi)]=0.82 V_0(\Delta\phi)f_0(\Delta\phi)+0.82 V_0(\pi-\Delta\phi)f_\pi(\Delta\phi)+\Re[U_{a}(\Delta\phi)][1-f_0(\Delta\phi)-f_\pi(\Delta\phi)]$.
In the figure we depict as well the analytic expressions $V_a(\Delta\phi)$~(solid curve) and $\Re(U_a(\Delta\phi))$~(bold solid curve).
}
\label{fig:1}
\end{figure}

\begin{figure}[t]
\includegraphics[width=\columnwidth]{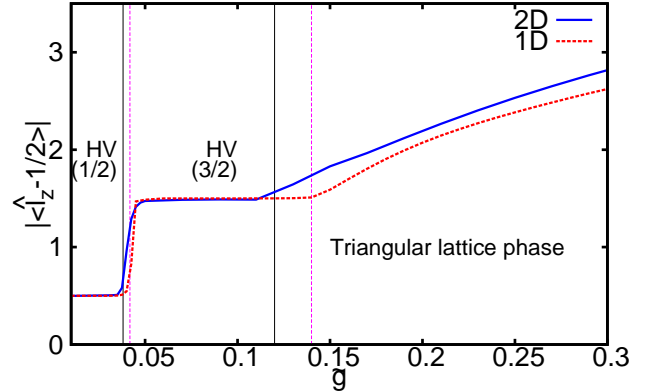}
\caption{(Color online) Comparison of the results for the average angular momentum  $|\langle \hat l_z-1/2 \rangle |$ versus $\tilde g$ for $\tilde\kappa=20$ between
the 2D model~(solid) and the effective 1D ring model~(dashed). The transitions from HV(1/2) to HV(3/2) and HV(3/2) to triangular lattice phases are represented by black solid vertical lines for the 2D model and pink dashed lines for 
the 1D model.}
\label{fig:2}
\end{figure}

We may then re-write:
\begin{equation}
E_{\mathrm{I}}=\frac{g\tilde\kappa^2}{4l_0^2}
\sum_{l_1,l_2,l_3,l_4}
a_{l_1}^*a_{l_2}a_{l_3}^*a_{l_4}\delta_{l_2+l_4,l_1+l_3} f_{l_1,l_2}^{l_3,l_4},
\label{eq:HI2}
\end{equation}
with
\begin{eqnarray}
f_{l_1,l_2}^{l_3,l_4}&\equiv&\int_0^\infty s ds\, e^{-2s^2} \left [ J_{l_1-1}J_{l_2-1} + J_{l_1}J_{l_2} \right ] \nonumber \\
&& \left [ J_{l_3-1}J_{l_4-1} + J_{l_3}J_{l_4} \right ],
\end{eqnarray}
where we use the simplified notation $J_m=J_m(\tilde\kappa s)$. Substituting $a_l=\int \frac{d\phi}{\sqrt{2\pi}}G(\phi)e^{-\ii l\phi}$ into Eq.~\eqref{eq:HI2} we obtain
after straightforward manipulations:
\begin{eqnarray}
\!\!\!\!\frac{E_{\mathrm{I}}}{\hbar\omega}\!&=&\! \frac{\tilde g\tilde\kappa^2}{16\pi^2}\! \int\! d\phi_1d\phi_2d\phi_3d\phi_4 G(\phi_1)^* G(\phi_2) G(\phi_3)^* G(\phi_4) \nonumber \\
&& W(\phi_1,\phi_2,\phi_3,\phi_4) A(\phi_1,\phi_2,\phi_3,\phi_4),
\label{eq:HI-tot}
\end{eqnarray}
with $\tilde g=g/(l_0^2\hbar\omega)$. In Eq.~\eqref{eq:HI-tot}, the function
\begin{eqnarray}
\!\!\!\!\! && A(\phi_1,\phi_2,\phi_3,\phi_4)= e^{\ii \Phi/2} \nonumber \\
\!\!\!\!&&\left [
\cos\left ( \frac{\phi_1-\phi_2}{2} \right )\cos\left ( \frac{\phi_3-\phi_4}{2} \right ) \right\delimiter 0 \nonumber \\
\!\!\!\!\!&+&
\left\delimiter 0 \cos\left ( \frac{\phi_1-\phi_4}{2} \right )\cos\left ( \frac{\phi_3-\phi_2}{2} \right )
\right ],
\end{eqnarray}
with $\Phi\equiv\phi_1-\phi_2+\phi_3-\phi_4$, stems from the particular form
of $\boldsymbol\eta_-(\phi)$.  The form of the function $A$ is hence specific to spin-$1/2$ BECs with spin-independent interactions. As we
show below $A$ is different for spinor BECs with higher spins and/or spin-dependent interactions. In contrast,
\begin{eqnarray}
&&W(\phi_1,\phi_2,\phi_3,\phi_4)=2\sum_{l_1,l_2,l_3,l_4}  \delta_{l_2+l_4,l_1+l_3}
  \nonumber \\
 && e^{\ii (l_1\phi_1+l_3\phi_3-l_2\phi_2-l_4\phi_4)}  \int s ds e^{-2s^2} J_{l_1}J_{l_2}J_{l_3}J_{l_4}
 \label{eq:W-1}
\end{eqnarray}
is a general function associated to the Rasba ring, valid for spinor BECs with arbitrary spin and with spin-dependent interactions,
as shown in Sec.~\ref{sec:RingModelSpin1}.
Interestingly, the function $W$ may be reduced to a closed analytical form~(see App.~\ref{sec:AppendixA}):
\begin{eqnarray}
&& W(\phi_1,\dots,\phi_4)=e^{-\frac{\tilde\kappa^2}{2}\cos^2(\Phi/4) \left [ \cos \left ( \frac{\phi_3-\phi_1}{2}\right )-\cos\left ( \frac{\phi_2-\phi_4}{2} \right ) \right ]^2} \nonumber \\
&& e^{-\frac{\tilde\kappa^2}{2}\sin^2(\Phi/4) \left [ \cos \left ( \frac{\phi_3-\phi_1}{2}\right )+\cos\left ( \frac{\phi_2-\phi_4}{2} \right ) \right ]^2} .
\label{eq:W}
\end{eqnarray}
For large $\tilde\kappa$, we may use the limit definition, $\lim_{\epsilon\rightarrow 0} \frac{e^{- x^2/4\epsilon}}{2\sqrt{\pi\epsilon}}=\delta(x)$, to obtain:
\begin{eqnarray}
&&\!\!\!\!\!\! W(\phi_1,\dots,\phi_4)=\frac{2\pi}{\tilde \kappa^2}
\nonumber \\
&& \!\!\!\!\!\!\delta\! \left [
\cos\!\left (\frac{\Phi}{4} \right )\!\! \left [ \cos\!\left ( \frac{\phi_3-\phi_1}{2}\right )\!-\!\cos\!\left ( \frac{\phi_2-\phi_4}{2} \right ) \right ]
\right ] \nonumber \\
&& \!\!\!\!\!\!\delta \left [
\sin\!\left (\frac{\Phi}{4} \right )\!\! \left [ \cos \!\left ( \frac{\phi_3-\phi_1}{2}\right )\!+\!\cos\!\left ( \frac{\phi_2-\phi_4}{2} \right ) \right ]
\right ].
\end{eqnarray}

\subsection{Interaction channels}
\label{subsec:Channels}

The function $W$ can be viewed as an approximate momentum conservation on the ring, which selects two interaction channels:
\begin{itemize}
\item type-(i) interactions:  $\phi_1\simeq \phi_2$ and $\phi_3\simeq \phi_4$, or $\phi_1\simeq \phi_4$ and $\phi_3\simeq \phi_2$;
\item type-(ii) interactions:  $\phi_3\simeq \phi_1+\pi$ and $\phi_4\simeq \phi_2+\pi$ ~(modulo $2\pi$).
\end{itemize}
For type-(i) interactions, $\phi_2\simeq \phi_1$ and $\phi_4\simeq \phi_3$, we may re-write
\begin{equation}
W(\phi_1,\phi_2,\phi_3,\phi_4)\simeq \frac{8\pi}{\tilde \kappa^2} \frac{\delta(\phi_4-\phi_3)\delta(\phi_2-\phi_1)}{|\sin(\phi_3-\phi_1)|},
\label{eq:W1}
\end{equation}
where the validity of the expression demands $\sin^2((\phi_3-\phi_1)/2) \gg 2/\tilde\kappa^2$.
For type-(ii) interactions, $\phi_3\simeq \phi_1+\pi$ and $\phi_4\simeq \phi_2+\pi$,  one obtains
\begin{equation}
W(\phi_1,\phi_2,\phi_3,\phi_4)\simeq \frac{8\pi}{\tilde \kappa^2} \frac{\delta(\phi_3-\phi_1-\pi)\delta(\phi_4-\phi_2-\pi)}{|\sin(\phi_2-\phi_1)|},
\label{eq:W2}
\end{equation}
for $\sin^2((\phi_2-\phi_1)/2) \gg 2/\tilde\kappa^2$.
Although these two types of effective interactions have been discussed in the context of homogeneous (i.e. untrapped) BEC with SOC~\cite{Gopalakrishnan2011,Zhou2013},
 their functional form is crucially different in the presence of confinement, especially due to the appearance of the sine function in the denominator
of the expressions above. Note that  this sine function in the denominators is problematic when it approaches zero.
We address this issue below.


\subsection{Effective interaction Hamiltonian}
\label{subsec:EffectiveHamiltonian}

Substituting the expressions for the $W$ function in Eq.~\eqref{eq:HI-tot} we obtain a simplified form of the interaction Hamiltonian:
\begin{widetext}
\begin{equation}
\frac{E_{\mathrm{I}}}{\hbar\omega}= \frac{\tilde g}{2} \int_0^{2\pi} d\phi \int_0^{2\pi} d\phi' V(\phi-\phi') |G(\phi)|^2|G(\phi')|^2
+ \frac{\tilde g}{2} \int_0^{\pi} d\phi \int_0^{\pi} d\phi'
U(\phi-\phi') G(\phi)^* G(\phi+\pi)^* G(\phi') G(\phi'+\pi).
\label{eq:HI-sim}
\end{equation}
\end{widetext}

The first term in $E_{\mathrm{I}}$ corresponds to type-(i) interactions, which hence may be understood as an effective
``long-range'' interaction in momentum space. The strength of the ``long-range'' interaction, given by the $V(\Delta\phi)$ function,
depends non-trivially on the angular separation $\Delta\phi=\phi-\phi'$, as discussed below.
The second term in $E_{\mathrm{I}}$ stems from the type-(ii) interactions, which
are characterized by the destruction of a pair of particles with opposite momenta, and the creation of another pair
of opposite momenta. The strength of the pair destruction-pair creation, $U(\Delta\phi)$, depends on the angular separation between the pairs.
In the following we discuss the form of the interaction potentials $V$ and $U$.

The general form of $V$ and $U$ is complicated. For $\Delta\phi$ sufficiently away from $0$ and $\pi$, we can provide a good approximation by using \eqref{eq:W1} and \eqref{eq:W2} to obtain
\begin{eqnarray}
V_{\mathrm a}(\Delta\phi) & \equiv & \frac{1+\cos^2\left ( \Delta\phi/2 \right )}{\pi |\sin(\Delta\phi)|},\label{eq:Va}\\
U_{\mathrm a}(\Delta\phi) & \equiv & \frac{2\cos (\Delta\phi)}{\pi|\sin(\Delta\phi)|}e^{\ii \Delta\phi}.\label{eq:Ua}
\end{eqnarray}
Note that these expressions are independent of $\tilde\kappa$. Interestingly, these expressions are identical to those found in homogeneous BECs~\cite{Gopalakrishnan2011,Zhou2013}
except for the crucial presence of the sine function in the denominator. On the other hand, in the vicinity of $\Delta\phi = 0$ or $\pi$, the approximation leading to \eqref{eq:W1} and \eqref{eq:W2} break down. To calculate $V$ and $U$ for all $\Delta\phi$ we can introduce a patching function $V_0$ and $U_0$, and express
\begin{eqnarray}
V(\Delta\phi) & = &
\begin{cases}
V_{a}\left(\Delta\phi\right) & |\Delta\phi-\pi/2|<.2\\
V_{0}\left(\Delta\phi\right) & {\rm otherwise}
\end{cases}
,\label{eq:V-patch}\\
U(\Delta\phi) & = &
\begin{cases}
U_{a}\left(\Delta\phi\right) & |\Delta\phi-\pi/2|<.2\\
U_{0}\left(\Delta\phi\right) & {\rm otherwise}
\end{cases}
.\label{eq:U-patch}
\end{eqnarray}

In order to evaluate the function $V_0$ close to $\Delta\phi=0$ we will need to use a series expansion. Note that the function $V(\Delta\phi)$ must be symmetric around $\Delta\phi=0$, and hence in the vicinity of $\Delta\phi=0$, it may be expanded in the form $V(\Delta\phi)\simeq V_0(\Delta\phi)\equiv \sum_{j=0}^\infty v_j |\Delta\phi|^j$. We can then assume the angular dependence is a Gaussian wavefunction
\begin{equation}
G(\phi)=f_G(\phi)\equiv \frac{e^{-\phi^2/2\delta\phi^2}}{\pi^{1/4}\sqrt{\delta\phi}},
\label{eq:Gtry}
\end{equation}
that is localized with a with a small width $\delta\phi\ll\pi$. For this particular angular wavefunction, only type-(i) interactions contribute, due to the absence of a wavefunction at opposite momenta.
The interaction energy~\eqref{eq:HI-sim} for the single Gaussian~\eqref{eq:Gtry} can be calculated analytically, and only $V_0$ contributes to give
$$
\frac{E_{\mathrm {int}}^{\mathrm {1G}}}{\hbar\omega}= \frac{\tilde g}{2}\sum_j \left [ \frac{\Gamma((j+1)/2)\delta\phi^j}{\sqrt{\pi}}\right ]v_j.
$$
The expansion coefficients $v_j$ are found by performing this calculation for a given $\delta\phi$, and equating the result with the energy that found from using $G(\phi)$ function using Eq.~\eqref{eq:HI}. Repeating this procedure for a range of $\delta\phi$, all relevant $v_j$ can be found.

To calculate $U_0$, we must repeat this procedure with a $G(\phi)$ formed by two non-overlapping Gaussians~(with total normalization $1$) of width $\delta\phi\ll \pi$, placed at $\pm\pi/2$. The interaction energy is of the form:
$$\frac{E_{\mathrm{int}}^{\mathrm {1G}}}{\hbar\omega}=\frac{1}{2}\frac{E_{\mathrm{int}}^{\mathrm {1G}}}{\hbar\omega}+\frac{\tilde g}{4} \sum_j \left [ \frac{\Gamma((j+1)/2)\Delta\phi^j}{\sqrt{\pi}}\right ](\tilde v_j + u_j/2),$$
where the
$V$ function in the vicinity of $\Delta\phi=\pi$ may be approximated by $V(\Delta\phi)\simeq \sum_{j=0}^\infty \tilde v_j |(\Delta\phi-\pi)|^j$, and
 in the vicinity of $\phi=0$, $\Re[U(\Delta\phi)]\simeq\sum_{j=0}^\infty u_j |\Delta\phi|^j$. We have numerically checked that $E_{\mathrm{int}}^{\mathrm{2G}}=E_{\mathrm{int}}^{\mathrm{1G}}$, and hence
 $v_j=\tilde v_j + u_j/2$. Finally, note that $\Re[U]$ must be symmetric around $\pi/2$, and hence the behavior at $\Delta\phi\simeq \pi$ is the same as that at $\Delta\phi\simeq 0$.

By properly matching the analytical expressions and the values in the vicinity of $\Delta\phi=0,\pi$, we obtain the final form of the $U$ and $V$ functions.
Taking $\tilde v_j=0.59 v_j$ and $u_j=0.82 v_j$, we have obtained for different $\tilde\kappa\gg 1$ values that the $V$ and $U$ functions calculated at $\Delta\phi\simeq 0$ and $\pi$ smoothly connect
with the analytical expressions~\eqref{eq:Va} and~\eqref{eq:Ua}. The exact value of the coefficients $v_j$, and hence the form of $V_0(\Delta\phi)$, depends however on $\tilde\kappa$.


\subsection{Effective one-dimensional Gross-Pitaevskii equation}
\label{subsec:1DGPE}

Employing Eqs.~\eqref{eq:HSOCT-sim} and~\eqref{eq:HI-sim} we may derive the effective one-dimensional GPE.
For $0<\phi<\pi$:
\begin{eqnarray}
&&i\frac{\partial}{\partial\tau} G(\phi,\tau)=\frac{1}{2\tilde\kappa^2}\left ( \hat l_z- \frac{1}{2}\right )^2G(\phi,\tau) \nonumber\\
&&+\tilde g \int_0^{2\pi} \!\!d\phi' V(\phi-\phi') |G(\phi',\tau)|^2 G(\phi,\tau)  \label{eq:1DGPE} \\
&&+\frac{\tilde g}{2} \int_0^\pi \!\!d\phi' U(\phi-\phi')G(\phi+\pi,\tau)^* G(\phi',\tau)G(\phi'+\pi,\tau), \nonumber
\label{eq:GPE1D}
\end{eqnarray}
with $\tau=\omega t$.
For $\phi>\pi$ the last line of the previous expression must be changed into
$\frac{\tilde g}{2} \int_\pi^{2\pi} U(\phi-\phi')G(\phi-\pi,\tau)^* G(\phi',\tau)G(\phi'-\pi,\tau)$.


\subsection{Understanding the phase diagram of a 2D BEC with SOC using the effective ring model}
\label{subsec:UnderstandingSpin12}

Figure~\ref{fig:1} shows the form of the $V$ and $U$ functions for $\tilde\kappa=20$. Note that the function $V$ is characterized by
the appearance of a local non-zero minimum at $\Delta\phi\simeq 0.6\pi$. The function $U$ presents a zero minimum at $\Delta\phi=\pi/2$.
Note that this peculiar dependence of the interaction strengths $U$ and $V$ stems from the $1/|\sin(\Delta\phi)|$ dependence of the $V_a$ and $U_a$ functions.
This dependence is characteristic of trapped condensates with SOC with $\hbar\omega$ much larger than the interaction energy, being absent in homogeneous BECs~\cite{Gopalakrishnan2011,Zhou2013}.
As we discuss in the following, the $1/|\sin(\Delta\phi)|$ dependence is crucial to understand the ground-state phases of trapped 2D BECs with an isotropic SOC,
and in particular the appearance of skyrmion lattice phases~\cite{Sinha2011,Hu2012,Xu2012}, whose origin remained up to now unclear.

\begin{figure}
\includegraphics[width=\columnwidth]{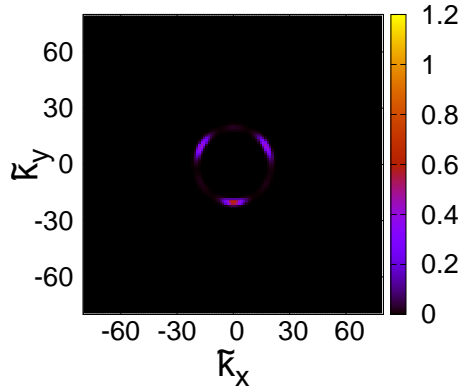}
\includegraphics[width=\columnwidth]{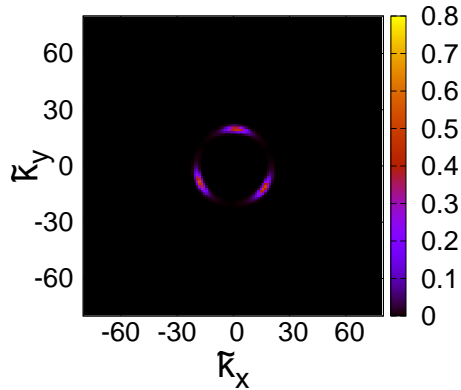}
\caption{(Color online) Momentum distribution~(as a function $\tilde k_x\equiv k_x l_0$ and $\tilde k_y\equiv k_y l_0$) of a spin-$1/2$ BEC with isotropic SOC for $\tilde\kappa=20$ and $\tilde g=0.19$ obtained from a direct numerical simulation of
Eq.~\eqref{eq:GPE2D} (top) and of the effective ring model~(bottom). A very similar triangular momentum distribution is observed with both model. }
\label{fig:3}
\end{figure}

\begin{figure}
\includegraphics[width=\columnwidth]{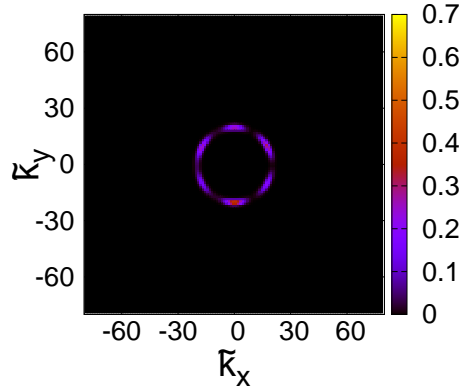}
\includegraphics[width=\columnwidth]{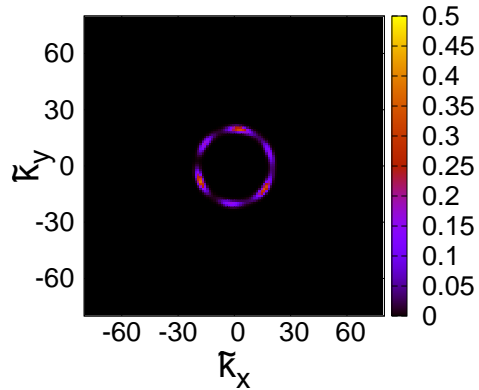}
\caption{(Color online) Same as Fig.~\ref{fig:3} but under different initial conditions for the imaginary time evolution. A very similar hexagonal
pattern appears. The energy of the triangular pattern of Fig.~\ref{fig:3} and of the hexagonal pattern of this figure is within our numerical accuracy basically identical~(see discussion in the main text).}
\label{fig:4}
\end{figure}

For vanishing interactions, it is clear from the form of $E_{\mathrm{SOC}}+E_{\mathrm{T}}$ that the lowest energy is given by the HV(1/2) phase, which has angular momentum $l=0$
or $1$~(we employ in the following the notation of Ref.~\cite{Sinha2011}). Note that the contribution of type-(i) interactions to the interaction energy of both HV(1/2) and HV(3/2) phases is identical. The HV(1/2) to HV(3/2) transition is hence given by the type-(ii) interactions. The transition occurs when
$\frac{\tilde g}{2}\int d\phi \int d\phi' \frac{1}{(2\pi)^2} U(\phi-\phi') [1-e^{-\ii 4(\phi-\phi')}]=\tilde\kappa^{-2}$.
Using the expression calculated above, one obtains for $\tilde\kappa=20$ that the HV(1/2) to HV(3/2) transition occurs at $\tilde g=2.34 (2\pi/\tilde\kappa^2)$, in excellent agreement
with the exact result, $2.35(2\pi/ \tilde\kappa^2)$, obtained from the direct imaginary time evolution of the 2D Gross-Pitaevskii equation~\cite{Sinha2011} .

The transition to the lattice phases results from the form of the $V$ function. Recall that in the homogeneous case, the interaction energy is clearly minimized by
placing the BEC in a plane-wave phase~(single momentum peak) or two opposite peaks~(stripe phase)~\cite{Wang2010}. However, the presence of a local minimum of the interaction energy
(which we stress is induced by the external trapping) allows, at intermediate interaction values of $\tilde g$, the system to minimize the energy by creating a lattice characterized by regular peaks in momentum space separated by an angle $\Delta\phi=2\pi/n$~\cite{Xu2012}. This solution has a large interaction energy but a smaller kinetic energy than the plane-wave or
stripe solution~(in the following we denote as ``kinetic energy'' the contribution of the $E_{\mathrm{SOC}}+E_{\mathrm{T}}$ term, which depends on the curvature, $\partial^2_\phi G$, of the angular distribution). Moreover, note that a lattice formed by three peaks
in momentum space may be approximated by three Gaussians placed at a separation of $2\pi/3$, i.e. very close to the minimum of $V$.

Note as well that in the $3$-peak case type-(ii) interactions are obviously irrelevant, since there are no pairs of opposite momenta.
In contrast, a $4$-peak square configuration given by Gaussian-like peaks with an angle separation of $\pi/2$~(which would lead to a square lattice phase) may present in principle type-(ii) interactions.
Note, however, that $U(\pi/2)=0$, i.e. quantum interference
results in the cancellation of type-(ii) processes also for the square configuration. This is a peculiar feature of spin-$1/2$ BECs, absent in spin-$1$ BECs, as discussed in Sec.~\ref{sec:RingModelSpin1}.
The absence of type-(ii) processes is crucial  in spin-$1/2$ BECs for the selection of the triangular phase against the square phase. Both phases have a similar interaction energy, but the square phase is characterized by momentum peaks with a narrower angular spreading, and hence by a larger kinetic energy.

It is interesting to comment on the case of an hexagonal phase, characterized by six momentum peaks along the Rashba ring formed by
three pairs of opposite Gaussians separated by an angle $\pi/3$.
We may compare the case of six Gaussians separated by an angle $\pi/3$ against the case of three Gaussians separated by $2\pi/3$, assuming in both cases Gaussians
of the same width. It is clear that the type-(i) part of the interaction energy is larger for the hexagonal case. However, for the $6$-peak case the type-(ii) interactions do not vanish.
The latter is crucial, since the interaction energy of the hexagonal phase may be then reduced by properly setting the phases of the Gaussian pairs such that the type-(ii)
contribution is negative~(the overall interaction energy is of course still repulsive).
 In particular for three pairs of opposite momenta with phases $\pi/3$, $0$ and $-\pi/3$, we have numerically checked that the type-(ii) contribution exactly cancels the increase of type-(i) interaction energy, and hence that the energy for $6$ and for $3$ non-overlapping Gaussians is the same (within our numerical accuracy). This degeneracy explains
the results obtained in Ref.~\cite{Sinha2011}.

Hence, the ring model allows for an intuitive understanding of the qualitative features observed in a spin-$1/2$ trapped BEC with dominant isotopic SOC in the weakly
interacting regime. Moreover, the effective 1D Hamiltonian given by Eqs.~\eqref{eq:HSOCT-sim} and~\eqref{eq:HI-sim} leads to a good quantitative agreement with the exact 2D result obtained from Eq.~\eqref{eq:GPE2D}. For example, in Fig.~\ref{fig:2} we compare the angular momentum
$|\langle \hat l_z-1/2 \rangle |$ in the 1D effective model and the 2D exact equation at $\tilde\kappa=20$. Clearly the 1D model recovers both the HV(1/2)-HV(3/2) transition and the HV(3/2)-lattice phase transition.
Moreover, although the ring model is not applicable for large interactions, the form of the $U$ and $V$ functions suggests that for sufficiently large $\tilde g$
the system should experience a first-order phase transition into the stripe~(or plane-wave) phase, as observed in the numerical simulations of the 2D Gross-Pitaevskii equation~\cite{Sinha2011}.
Note that this occurs when the system jumps from the local interaction minimum  to the global one at $\Delta\phi=0,\pi$, since the larger
kinetic energy is eventually compensated by the smaller interaction energy for a sufficiently large $\tilde g$.



\section{Ring model for spin-1 condensates}
\label{sec:RingModelSpin1}

In the previous section we have obtained an effective ring model for the specific case of a spin-$1/2$ condensate. The procedure is, however, general for spinor condensates
of any spin and arbitrary, possibly spin-dependent, short-range interactions, in the regime of dominant SOC and weak interactions~($\hbar\omega$ much larger than the interaction energy).
In this section we illustrate the use of the general method for a more complicated system, namely a spin-$1$ condensate with spin-dependent interactions.


\subsection{Effective ring model}
\label{subsec:RingModel}

We now consider the case of $F=1$.
The lowest branch, again with eigenenergy $\epsilon_-=\frac{\hbar^2}{2m}(k-\kappa)^2$, is characterized by the eigenvector
$$\boldsymbol\eta_-(\phi)=\frac{1}{2}
   \begin{pmatrix}
 e^{-\ii\phi}\\ -\sqrt{2}\\ e^{\ii\phi} \\
        \end{pmatrix}
,$$ where the different entries of the vector correspond to the
Zeeman components $m=-1,0,1$. As in the spin-$1/2$ case, we project into the lowest band,
obtaining the same expression~\eqref{eq:Psi}, but with the eigenvector $\boldsymbol\eta_-(\phi)$ of the spin-$1$ case.
The non-interacting part of the density functional acquires the form:
\begin{equation}
E_{\mathrm{SOC}}+E_\mathrm{T}=\left (\frac{m\omega^2}{2\kappa^2} \right ) \int d\phi G(\phi)^* \hat l_z^2 G(\phi).
\end{equation}
Transforming $G(\phi)=\sum_l a_l e^{\ii l\phi}/\sqrt{2\pi}$, and assuming that only angular momenta $l\ll \tilde\kappa$ contribute to $G(\phi)$,
we obtain the form of the spinor in coordinate space:
\begin{equation}
 \boldsymbol\Psi(\vec{r}) = \frac{\sqrt{\tilde\kappa}}{2\pi^{1/4}}e^{-s^{2}/2}\sum_l a_{l}
  e^{\ii l\alpha}\!\!
   \begin{pmatrix}
     \ii^{l-1}e^{-\ii\alpha}J_{l-1}(\tilde\kappa s) \\
     -\sqrt{2}\ii^{l}J_{l}(\tilde\kappa s) \\
     \ii^{l+1}e^{\ii\alpha}J_{l+1}(\tilde\kappa s) \\
   \end{pmatrix}.
   \label{eq:spinor1}
\end{equation}

Contrary to the case of spin-$1/2$ condensates, the ground-state of the non-interacting spin-$1$ BEC is unique and
given by $a_0=1$, $a_{l>0}=0$. From Eq.~\eqref{eq:spinor1} one sees that the non-interacting ground-state is characterized by
counter-propagating vortices in $m=\pm 1$ and a vortex-less $m=0$ component (HV(0) phase).

\begin{figure}[t]
\includegraphics[width=\columnwidth]{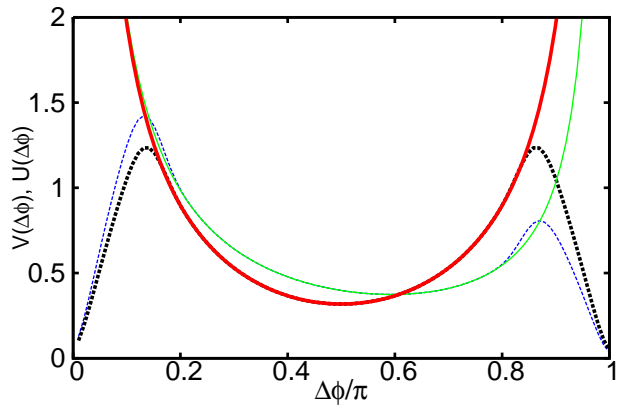}
\caption{(color online) Functions $V(\Delta\phi)$~(dashed curve) and $U(\Delta\phi)$~(bold dashed curve) for a spin-$1$ BEC with $\tilde\kappa=20$. In the figure we have depicted as well the
analytical expressions $V_a(\Delta\phi)$~(solid curve) and $U_a(\Delta\phi)$~(bold solid curve).}
\label{fig:5}
\end{figure}

The general form of the interacting part of the energy functional of a spin-$1$ spinor condensate is of the form~\cite{Ho1998,Ohmi1998}:
\begin{eqnarray}
E_{\mathrm I}&=&\!\!\int d^3 r \Big \{ \left ( \frac{g_0+2g_2}{6}\right ) |\psi_0|^4+ \frac{g_2}{2}\left [ |\psi_1|^4 +|\psi_{-1}|^4 \right ] \nonumber \\
&+&\!\!\left ( \frac{g_2+2g_0}{3}\right )|\psi_1|^2|\psi_{-1}|^2 + g_2 \left ( |\psi_1|^2+|\psi_{-1}|^2 \right ) |\psi_0|^2 \nonumber \\
&+&\!\!\left ( \frac{g_2-g_0}{3}\right )\left [ \psi_1^*\psi_{-1}^* (\psi_0)^2+\mathrm {c.c.} \right ] \Big\},
\end{eqnarray}
where $g_{\mathrm S}=4\pi\hbar^2 a_{\mathrm{sc}}(S)/m$, with  $a_{\mathrm{sc}}(S)$ the $s$-wave scattering length for the
channel of total spin $S=0$ and $2$.  When writing $E_\mathrm{I}$ above we have
assumed that the form of the interactions is not modified by the spin-orbit fields, such as in the case of magnetically generated spin-orbit coupling~\cite{Anderson2013,Xu2013}.
Employing expression~\eqref{eq:Psi}, but with the eigenvector $\boldsymbol\eta_-(\phi)$ of the spin-$1$ case, we obtain
again expression~\eqref{eq:HI-tot}, with $\tilde g=g_0/(l_0^2\hbar\omega)$, but with a different function
\begin{eqnarray}
&&A(\phi_1,\phi_2,\phi_3,\phi_4)=\left ( \frac{1+2\chi}{6}\right ) \nonumber \\
&+&\frac{\chi}{4}\cos(\phi_1+\phi_3-\phi_2-\phi_4) \nonumber \\
&+& \left ( \frac{\chi+2}{12}\right ) \cos(\phi_1-\phi_3)\cos(\phi_2-\phi_4) \nonumber \\
&+&\!\!\left ( \frac{\chi-1}{6} \right ) \left [ \cos(\phi_1-\phi_3)+\cos(\phi_2-\phi_4) \right ]  \nonumber \\
&+& \frac{\chi}{4} \Big [ \cos(\phi_1-\phi_4)+\cos(\phi_3-\phi_2) \nonumber \\
&+&\cos(\phi_3-\phi_4)+\cos(\phi_1-\phi_2) \Big ] ,
\end{eqnarray}
where $\chi=g_2/g_0$.
Since the $W$ function is the same as in the spin-$1/2$ case, we may employ Eqs.~\eqref{eq:W1} and~\eqref{eq:W2}, to obtain the corresponding $V$ and $U$ functions
sufficiently far from $\Delta\phi=0$ or $\pi$. For $\chi=1$, we obtain:
\begin{eqnarray}
V_a(\Delta\phi)&=&\frac{5+2\cos\Delta\phi+\cos^2\Delta\phi}{4\pi |\sin\Delta\phi|}, \label{eq:Vs1} \nonumber \\
U_a(\Delta\phi)&=&\frac{3+\cos 2\Delta\phi}{2\pi |\sin\Delta\phi|}, \label{eq:Us1}
\end{eqnarray}
Note, that as spin-$1/2$ BECs, the $V$ and $U$ functions are independent of $\tilde\kappa$ in the vicinity  of $\Delta\phi=\pi/2$.
In the vicinity of $\Delta\phi=0$ or $\pi$ we proceed as in the previous section to obtain $V_0(\Delta\phi)$ and $U_0(\Delta\phi)$, which as for the spin-$1/2$ case is $\tilde\kappa$ dependent.
In Fig.~\ref{fig:5} we show the form of the $V$ and $U$ functions for $\tilde\kappa=20$.


\subsection{Understanding the properties of spin-$1$ BECs with SOC using the ring model}
\label{subsec:UnderstandingSpin1}

\begin{figure}[t]
\includegraphics[width=\columnwidth]{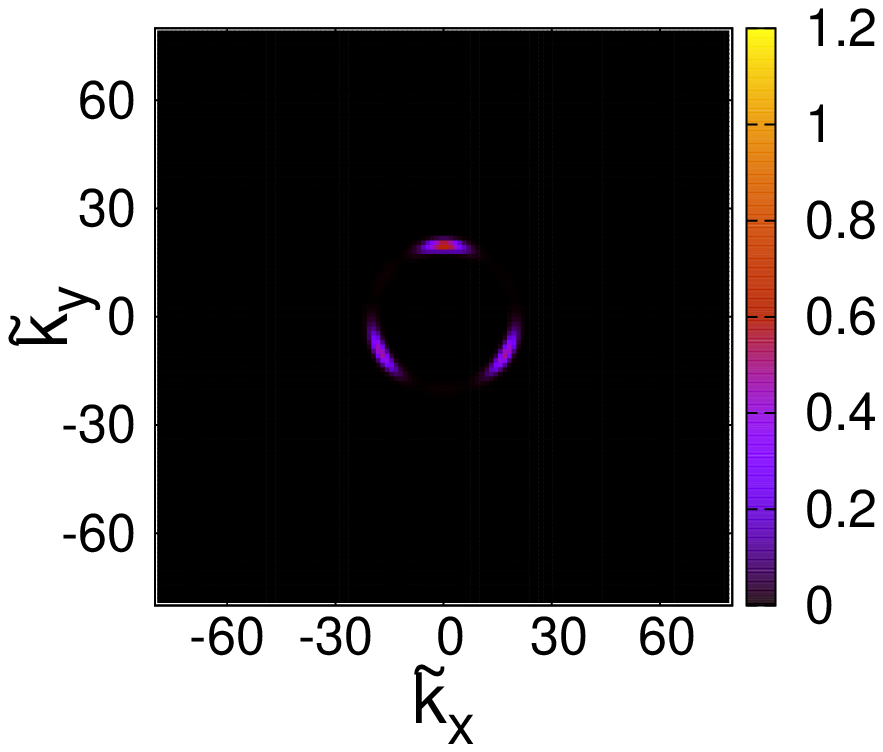}
\includegraphics[width=\columnwidth]{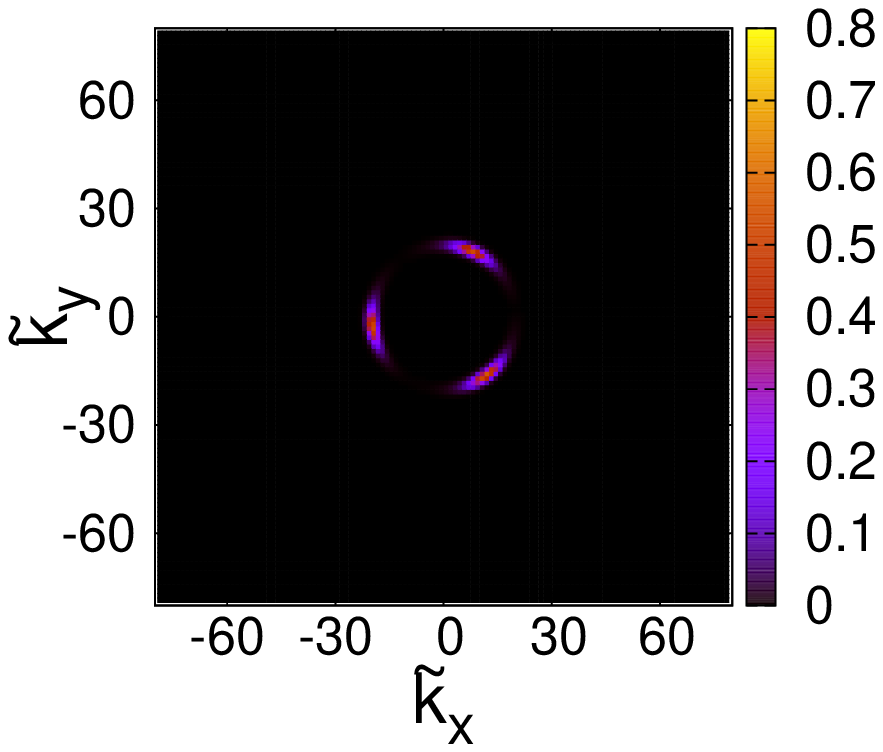}
\caption{(Color online) Momentum distribution~(as a function $\tilde k_x\equiv k_x l_0$ and $\tilde k_y\equiv k_y l_0$) of a spin-$1/2$ BEC with isotropic SOC for $\tilde\kappa=20$ and $\tilde g=0.2$ obtained from a direct simulation of
Eq.~\eqref{eq:GPE2D} (top) and of the effective ring model~(bottom). A very similar square momentum distribution~(triangular lattice phase) is observed in both cases.}
\label{fig:7}
\end{figure}

\begin{figure}[t]
\includegraphics[width=\columnwidth]{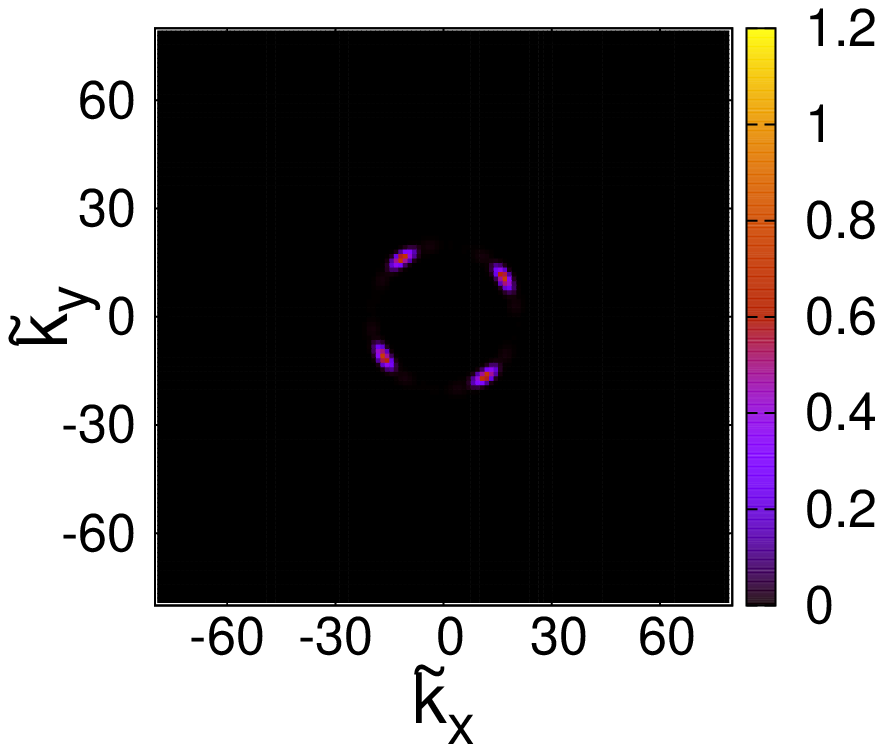}
\includegraphics[width=\columnwidth]{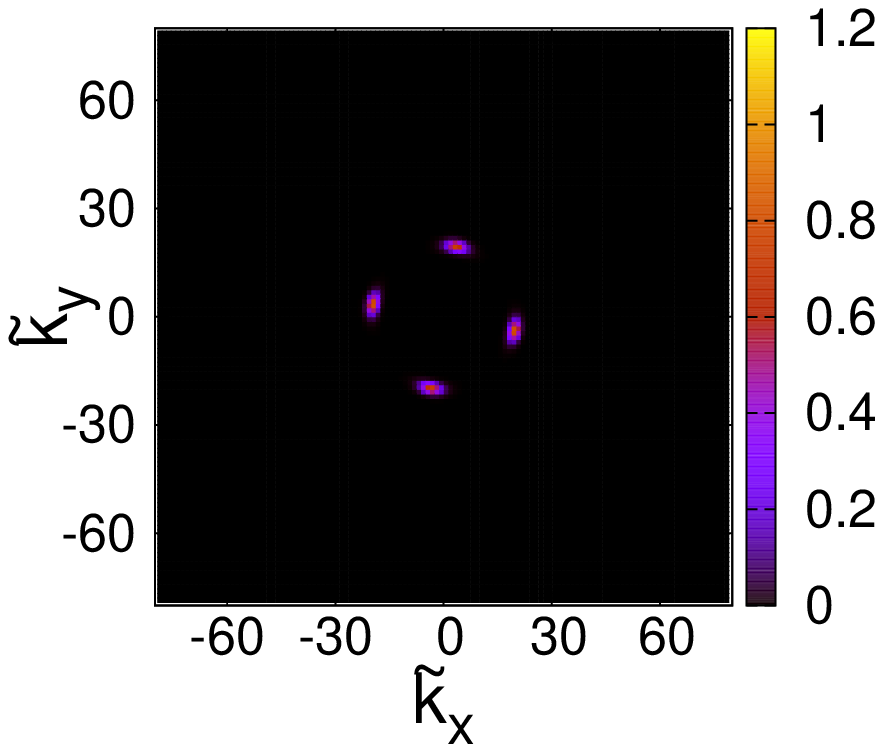}
\caption{(Color online) Momentum distribution~(as a function $\tilde k_x\equiv k_x l_0$ and $\tilde k_y\equiv k_y l_0$) of a spin-$1$ BEC with isotropic SOC for $\tilde\kappa=20$ and $\tilde g=0.8$ obtained from a direct simulation of
Eq.~\eqref{eq:GPE2D} (top) and of the effective ring model~(bottom). A very similar triangular momentum distribution~(square lattice phase) is observed in both cases.}
\label{fig:8}
\end{figure}

As for the spin-$1/2$ case, the effective ring model, and in particular the form of the functions $U$ and $V$ allows for an intuitive understanding of the
properties of spin-$1$ BECs under isotropic SOC. Figure~\ref{fig:4} compares, for $\tilde\kappa=20$, the expectation value $|\langle\hat l_z\rangle |$
obtained using directly the 2D GPE with the effective spin-1 1D GPE in Eq.~\eqref{eq:GPE1D}.
The direct solution of the 2D GPE shows that, as mentioned above,
the non-interacting BEC is in the HV(0) phase. As interactions are increased, the system experiences a phase transition into the HV(1) phase, characterized by $l_z=\pm 1$.
As for spin-$1/2$, for a sufficiently large $\tilde g$ the system enters into a triangular lattice phase, characterized by three peaks along the ring.
However contrary to the spin-$1/2$ BEC, there is a second phase transition into a square lattice for a large-enough $\tilde g$~\cite{Xu2012}. All these features are well reproduced by the ring model~(see Figs.~\ref{fig:6}, \ref{fig:7} and  \ref{fig:8}).

\begin{figure}[t]
\includegraphics[width=\columnwidth]{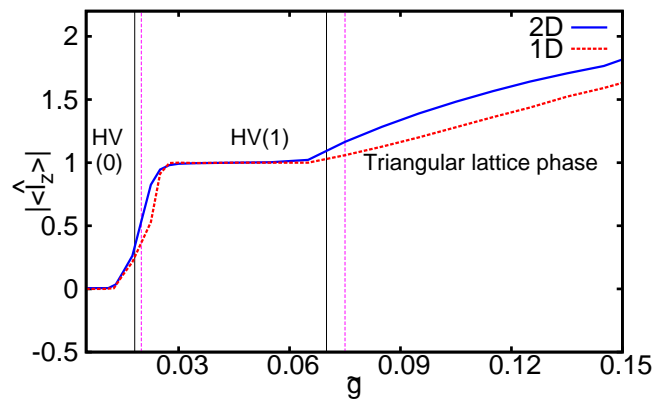}
\caption{(color online) Mean value of the angular momentum $|\langle \hat l_z \rangle |$ as a function of $\tilde g$ for a spin-$1$ BEC with $\tilde\kappa=20$. We compare the results
obtained from the 2D GPE~\eqref{eq:GPE2D} (solid) and from the ring model~(dashed). The transitions from HV(0) to HV(1) and from HV(1) to a triangular latice phase are represented by black solid vertical lines for the 2D model and pink dashed lines for the 1D model.}
\label{fig:61}
\end{figure}

The ring model also provides a clear insight on the physics behind the different lattice phases. To understand why the square phase
is preferred for sufficiently large $\tilde g$ it is crucial to realize that for the spin-$1$ case $U(\pi/2)=1/\pi$, whereas for the spin-$1/2$ case $U(\pi/2)=0$.
Note that, as mentioned above, the type-(ii) interactions depend on the phase of the $G(\phi)$ function.
It is hence possible to arrange the angular dependence of the phase such that the type-(ii) contribution to the interaction energy is minimized.
Assuming that the $G(\phi)$ function is formed by four separated narrow Gaussian-like wave packets, $f_G(\phi)$, at $j\pi/2$, with $j=0,1,2,3$,
$G(\phi)=\frac{1}{2} \sum_{j=0}^3 f_G(\phi-j\pi/2)e^{\ii \theta_j}$, we obtain that the interaction energy is proportional to $V(\pi/2)/2+\frac{1}{4}U(\pi/2)\cos\theta$, with
$\theta=\theta_0+\theta_2-\theta_1-\theta_3$. The energy is hence minimized for $\theta=\pi$, for which the contribution of the type-(ii) interactions is actually negative. A similar analysis for a triangular lattice results in an interaction energy proportional to $2V(2\pi/3)/3$.
Since for spin-$1$ BECs $U(\pi/2)$ is comparable to $V(\pi/2)$ and $V(2\pi/3)$, it is hence clear that the interaction energy of the square lattice may become significantly smaller
than that of the triangular lattice. This mechanism was crucially absent in the spin-$1/2$ case, since $U(\pi/2)=0$, and hence for spin-$1/2$ the triangular lattice was selected.
From Figs.~\ref{fig:7} an~\ref{fig:8}  it is however clear that the kinetic energy, being dependent on $\partial_\phi^2G$, is larger in the square lattice, explaining why there is an intermediate triangular lattice phase.

For $\tilde\kappa=20$~(the case of Figs.~\ref{fig:6}, \ref{fig:7} and  \ref{fig:8}) the direct numerical simulation of the 2D GPE
shows a triangular-to-square lattice phase transition at $\tilde g\simeq 0.28$, which is in very good quantitative agreement with the result obtained from the effective 1D ring model~($\tilde g\simeq 0.23$). We have also checked in our direct numerical simulation of the 2D GPE that the square lattice is characterized by $\theta=\pi$
as discussed above. Due to the minimization of the type-(ii) interactions induced by the relative phase arrangement the square lattice is very robust, and from our numerical simulation of the
2D GPE we observe that it remains the ground state for $\tilde g\gg 1$, well beyond the validity regime of the thin ring model.



\section{Outlook}
\label{sec:Outlook}

In this paper we have derived an effective quasi-one-dimensional ring model in momentum space for the study of two-dimensional BECs under dominant isotropic SOC and weak-enough interactions. The model, which may be generally applied to spinor BECs with arbitrary spin and spin-dependent interactions, reduces the BEC physics to the angular dependence along the Rashba ring. Two main energy contributions characterize this physics, the ``kinetic energy'' induced by the effective dispersion in momentum space introduced by the external trap, and
the interaction energy. The latter is provided by two types of interactions, an effective ``long-range'' interaction between two momentum components in the Rashba ring~(type-(i) interactions), and the destruction/creation of pairs of particles of opposite momentum in the ring~(type-(ii) interactions). Although these two types of interactions also occur naturally in the absence of trapping~\cite{Gopalakrishnan2011,Zhou2013}, we have shown that the presence of the trap introduces a peculiar angular dependence for these interactions, which is responsible for the appearance of skyrmion lattice phases in trapped BECs. We have shown that the ring model permits an intuitive understanding of the ground-state phases of condensates with isotropic SOC, well reproducing the qualitative and even quantitative features of the exact 2D model.

The ring model may be applied as well to systems with weakly anisotropic dispersion. This is in particular the case of realistic SOC implementations, that converge to an isotropic ring-like
dispersion only at large laser intensities~\cite{Campbell2011}. For large but finite intensities, the lowest-branch dispersion for a $4$ laser arrangement acquires the form
$$\epsilon_-(q,\phi)\simeq \frac{\hbar^2}{2m}(q-\kappa)^2 + A \cos2\phi,$$
where the constant $A$ scales inversely with the laser intensity~\cite{Campbell2011}.
The extra anisotropic term may be straightforwardly added to the non-interacting Hamiltonian, resulting in four energy minima along the Rashba ring separated by an angle $\pi/2$. At finite intensities and weak interactions the condensate will occupy these minima. The form of the interactions derived in this paper, however, will remain valid. In particular, the fact that $U(\pi/2)=0$~($>0$) in spin-$1/2$~(spin-$1$) BECs is expected to play a crucial role in the properties of BECs in these four-minima arrangements.

Finally, we would like to note that the ring model is interesting well beyond the description of the ground-state mean-field phases of BECs. It may be employed not only for the study of excitations and dynamics (employing, respectively, the effective quasi-one-dimensional Bogoliubov-de Gennes equations and the time-dependent GPE associated to Eq.~\eqref{eq:GPE1D}), but also for the study of beyond-mean-field physics, since the derivation of the ring model and the effective interactions does not rely on mean-field approximations: one could use Eq.~\ref{eq:HI-sim} with $G(\phi)$ replaced with a field operator $\hat{G}(\phi)$. The analysis of these problems will be the subject of further research.



\acknowledgements

We thank G. Juzeliunas, P. \"Ohberg, and M. Valiente for stimulating discussions. This work was supported by the Cluster of Excellence QUEST and the CAS-DAAD scholarship. B.M.A. acknowledge the financial support by the NSF through the Physics Frontier Center at JQI, and the ARO with funds from both the Atomtronics MURI and DARPA's OLE Program.

 \emph{Note added:} After the submission of this work, we became aware of a related work on few-electron dots\cite{Amin2014}.

\appendix

\section{Derivation of the $W$ function}
\label{sec:AppendixA}

In this appendix we present the derivation of the simplified form~\eqref{eq:W} of the $W$ function. We may re-write Eq.~\eqref{eq:W-1} in the form
\begin{widetext}
\begin{eqnarray}
W(\phi_{1},\phi_{2},\phi_{3},\phi_{4})& =& 2\sum_{L,l,l'}{e^{\ii L\Phi}
   e^{-\ii l(\phi_{1}-\phi_{2})}e^{\ii l'(\phi_{2}-\phi_{4})}}
   \int_0^\infty s\mathrm{d}s\, e^{-2 s^{2}}J_{L-l} J_{L-l'}J_{L+l}J_{L+l'} \nonumber  \\
 &+& 2\sum_{L,l,l'}e^{\ii (L+\frac{1}{2})\Phi}
   e^{-\ii (l+\frac{1}{2})(\phi_{1}-\phi_{3})}e^{\ii (l'+\frac{1}{2})(\phi_{2}-\phi_{4})}
   \int_0^\infty s\mathrm{d}s\, e^{-2 s^2}J_{L-l}J_{L-l'}J_{L+l+1}J_{L+l'+1},
\end{eqnarray}
\end{widetext}
with $J_l=J_L(\tilde\kappa s)$.
Employing the identities
  $2 \pi J_{L-l}J_{L+l} \!=\!(-1)^{L-l}\!\!\int\mathrm{d}u\, e^{-\ii 2Lu}J_{2l}(2\tilde\kappa s\cos{u})$ and
  $2 \pi J_{L-l}J_{L+l+1} \!=\!(-1)^{L-l}\!\!\int\mathrm{d}u\, e^{-\ii (2L+1)u}J_{2l+1} (2\tilde\kappa s\cos{u})$,
we obtain
\begin{widetext}
\begin{eqnarray}
 W(\phi_{1},\phi_{2},\phi_{3},\phi_{4}) &=&\frac{1}{2\pi^2}\int s\mathrm{d}s\, e^{-2 s^{2}}\int \mathrm{d}u \mathrm{d}\bar{u}
  \sum_{L,l,l'} (-1)^{l+l'}  \{e^{\ii L(\Phi-2(u+\bar{u}))} e^{-\ii l(\phi_{1}-\phi_{3})} e^{\ii l'(\phi_{2}-\phi_{4})}
 J_{2l}(x) J_{2l'}(\bar{x}) \nonumber \\
 &+&e^{\ii (L+\frac{1}{2})(\Phi-2(u+\bar{u}))} e^{-\ii (l+\frac{1}{2})(\phi_{1}-\phi_{3})} e^{\ii (n+\frac{1}{2})(\phi_{2}-\phi_{4})}
  J_{2l+1}(x)J_{2n+1}(\bar{x})\}
\end{eqnarray}
\end{widetext}
with $x=2\tilde\kappa s\cos{u}$ and $\bar{x}=2\tilde \kappa s \cos{\bar{u}}$. Using the identities
$\cos(x\cos{\eta})=\sum_{l}(-1)^{l}e^{-\ii 2l\eta}J_{2l}(x)$ and
$\sin(x\cos{\eta})=\sum_{l}(-1)^{l}e^{-\ii (2l+1)\eta}J_{2l+1}(x)$,
and employing $\theta=2(u-\bar{u})$ and $\alpha=\frac{u+\bar{u}}{2}$, we obtain
\begin{widetext}
\begin{eqnarray}
  &&W(\phi_{1},\phi_{2},\phi_{3},\phi_{4})=\frac{1}{4\pi}\! \int s\mathrm{d}s e^{-2s^{2}}\!\! \int_{-4\pi}^{4\pi} \!\mathrm{d}\theta
   \int_{-\pi}^{\pi}\!\mathrm{d}\alpha\sum_{n}\delta(\Phi-\theta+2\pi n) \nonumber \\
  &&\left \{\cos\left(x\cos \left ( \frac{\phi_{3}-\phi_{1}}{2}\right ) \right)
   \cos\left(\bar{x}\cos\left ( \frac{\phi_{4}-\phi_{2}}{2}\right )\right)+  e^{\ii \frac{\phi-\theta}{2}}\sin\left(x\cos\left ( \frac{\phi_{3}-\phi_{1}}{2}\right )\right)
   \sin\left(\bar{x}\cos\left ( \frac{\phi_{4}-\phi_{2}}{2}\right)\right ) \right \} \nonumber \\
  &&=\frac{2}{\pi}\int r\mathrm{d}re^{-2 r^{2}}\int \mathrm{d}\alpha \cos\left[2\kappa r
   \left(\cos\left(\alpha+\frac{\Phi}{4}\right)\cos\left ( \frac{\phi_{1}-\phi_{3}}{2}\right ) -
     \cos\left(\alpha-\frac{\Phi}{4}\right)\cos\left (\frac{\phi_{2}-\phi_{4}}{2}\right ) \right)\right ] .
     \end{eqnarray}
\end{widetext}
Solving the Gaussian integral leads to Eq.~\eqref{eq:W}.

\end{document}